\documentclass[nofootinbib]{revtex4}

\usepackage{graphicx}

\def\be{\begin{equation}}
\def\ee{\end{equation}}
\def\ve{\varepsilon}
\def\eps{\varepsilon}
\def\GeV{{\rm\ GeV}}
\def\CM{{\cal M}}

\def\Im{\mathop{\rm Im}\nolimits}

\begin{document}

\title{Two-photon exchange in elastic $e\pi$ scattering}

\author{Dmitry~Borisyuk}
\author{Alexander~Kobushkin}

\affiliation{Bogolyubov Institute for Theoretical Physics,
Metrologicheskaya street 14-B, 03680, Kiev, Ukraine}

\begin{abstract}
 We calculate two-photon exchange amplitude
 for the elastic electron-pion scattering
 in the dispersion-relation inspired approach,
 including both elastic and inelastic contributions.
 The latter was modelled as a sum of
 $\rho$ and $b_1(1235)$ meson contributions.
 We find that at $Q^2 \lesssim 2\GeV^2$ the elastic contribution is dominant,
 similarly to electron-proton scattering case.
 At higher $Q^2$ the inelastic contribution is not negligible,
 but still smaller than the elastic one.
 We also explain observed rapid amplitude growth at backward angles.
\end{abstract}

\maketitle

\section{Introduction}
 In last years, two-photon exchange (TPE)
 is widely discussed in the literature.
 The role of TPE in the elastic electron-proton scattering
 was studied most thoroughly.
 In particular, it was shown that TPE can be responsible
 for the discrepancy between Rosenbluth and polarization transfer
 measurements of the proton form factors (FFs) \cite{epTPE, BMTres}.
 These intriguing results have triggered a study of TPE effects
 in other processes, such as
 elastic $ed$ \cite{edTPE}, deep inelastic scattering \cite{DIS-TPE}, and so on.
 
 Two recent papers \cite{piTPE1,piTPE2} discuss TPE in the elastic $e\pi$ scattering.
 Both are limited to so-called elastic contribution,
 and report significant increase of TPE amplitude at backward angles.
 These papers use different approaches to loop integral calculation:
 in Ref.~\cite{piTPE1} it was calculated approximately,
 under the assumption that both photons carry about a half
 of the  transferred momentum, whereas in Ref.~\cite{piTPE2}
 the loop integral was expressed through
 't~Hooft-Veltman $n$-point functions.
 However, the starting expression for the TPE diagram
 in Refs.~\cite{piTPE1,piTPE2} is written somewhat heuristically.
 In particular the (virtual) Compton scattering amplitude,
 constructed in the same manner, might be not gauge-invariant%
\footnote{%
 For example, in Ref.~\cite{piTPE2} TPE amplitude results from contracting
 leptonic and hadronic tensors, $L_{\mu\nu}$ and $H_{\mu\nu}$.
 It is tempting to identify the hadronic tensor,
 after proper crossing symmetrization,
 with the virtual Compton scattering amplitude:
 $T_{\mu\nu} \sim H_{\mu\nu}+H_{\nu\mu}(k_1\leftrightarrow k_2)$.
 However with $H_{\mu\nu}$ from Eq.(8) of Ref.~\cite{piTPE2} we have
 $T_{\mu\nu}k_1^\nu = 2k_{1\mu}$ instead of 0, as required by gauge invariance.
}.
%(see also discussion in the Introduction of Ref.~\cite{ourDisp}).
 
 In the present paper we evaluate TPE amplitude for the elastic $e\pi$
 scattering using dispersion-relation inspired approach,
 developed in Ref.~\cite{ourDisp}. The positive features % advantages
 of this approach are
 (for more detail see discussion in the Introduction of Ref.~\cite{ourDisp}):
 \begin{itemize}
 \item no need for off-shell FFs,
 \item clear and unambiguous definition of elastic and inelastic contributions,
 \item correct analytic structure of the resulting TPE amplitude
  (i.e. it is an analytic function with only branch cuts dictated by unitarity).
 \end{itemize}
 We evaluate both elastic and inelastic contributions.
 Following common practice \cite{BMTres}, the latter
 is modelled as a sum of resonance contributions,
 namely, $\rho$ and $b_1(1235)$ meson contributions.
 Certainly, contributions of other meson resonances can be easily included.

\section{TPE amplitude}

 Throughout the paper we use the notation, similar to Ref.~\cite{ourDisp}.
 The initial and final pion (electron) momenta are denoted $p$ and $p'$
 ($k$ and $k'$), respectively. The transferred momentum is $q = p'-p$,
 the pion mass is $M$, the electron mass $m$
 is assumed to be infinitely small.

 In one-photon exchange approximation,
 the elastic electron-pion scattering amplitude is
\be \label{1ph}
 \CM_{fi} = - \frac{4\pi\alpha}{q^2} \bar u'\gamma^\mu u \, (p+p')_\mu \, F(q^2)
\ee
 where $\alpha$ is fine structure constant and
 the real-valued function $F(q^2)$ is called pion electromagnetic FF.
 Since the pion has zero spin, it is easy to see that even in general case
 the amplitude keeps the same structure (\ref{1ph}),
 if we neglect the electron mass.
 The only difference with one-photon exchange case is that
 the function $F$ becomes complex and depends on both $t \equiv q^2$
 and $\nu = (p+p')(k+k')$.
 We may write
\be
 F(t,\nu) = F(t) + \delta F(t,\nu) + O(\alpha^2),
\ee
 where $\delta F$ is TPE contribution. Because of charge conjugation
 and crossing symmetry, $\delta F$ should be an odd function of $\nu$.
 
 To calculate the TPE amplitude, we start with its absorptive part,
 for which we have the unitarity condition
\be \label{uni}
 \CM_{fi} - \CM_{if}^* = \frac{i}{4\pi^2} \sum_{h} \int
    \CM_{fn}\CM_{in}^* \theta(p''_0) \delta(p''^2-M_h^2)
    \theta(k''_0) \delta(k''^2) d^4k''
\ee
 where subscript $n$ denotes intermediate state,
 which consists of the electron with momentum $k''$
 and some hadronic state $h$.
 We restrict ourselves to two-particle intermediate states,
 thus $h$ can only be a single meson with positive charge
 and negative $C$-parity ($\pi$, $\rho$ an so on).
 Its mass is denoted $M_h$ and momentum is $p''=p+k-k''$. 
 Retaining only the term corresponding to $h=\pi$,
 we obtain so-called elastic contribution $\delta F^{\rm (el)}$.
% 
% (the r.h.s. is restricted to two-particle intermediate states, which can 
% be $e\pi$, $e\rho$ and so on). Retaining only the term corresponding to
% $n = e\pi$, we obtain so-called elastic contribution $\delta F^{\rm (el)}$.
 Other contributions are referred to as inelastic ones.

\subsection{Elastic contribution}\label{Sec:el}

 Substituting the elastic amplitude (\ref{1ph})
 in the unitarity condition (\ref{uni}), we obtain
\be
 \bar u'\gamma^\mu u \, (p+p')_\mu \, \Im \delta F^{\rm (el)} (t,\nu) =
 -\frac{\alpha t}{2\pi}
  \int \bar u'\gamma^\mu \hat k'' \gamma^\nu u \,
    (p'+p'')_\mu (p''+p)_\nu \, \bar F(t_1) \bar F(t_2)
    \theta(p''_0) \delta(p''^2-M^2) \theta(k''_0) \delta(k''^2) d^4k''
\ee
 where $t_1 = (p''-p)^2$, $t_2 = (p''-p')^2$ and $\bar F(t) = F(t)/t$,
 and, after some transformations
\be
 \Im \delta F^{\rm (el)} (t,\nu) = -\frac{\alpha t}{2\pi}(\nu-t)
  \int \left\{ 1 + \frac{2M^2+\nu-t}{\nu^2+t(4M^2-t)} t_p \right\}
    \bar F(t_1) \bar F(t_2)
    \theta(p''_0) \delta(p''^2-M^2) \theta(k''_0) \delta(k''^2) d^4k''
\ee
 where 
 $t_p = t_1+t_2-t$.
 
 Now, to obtain box-type amplitude $\delta F^{\rm (el)}_{\rm box}$
 we should first change, according to Ref.~\cite{ourDisp},
\be
 \theta(p''_0)\delta(p''^2-M^2) \theta(k''_0)\delta(k''^2)
 \to \frac{1}{2\pi^2 i}\frac{1}{k''^2(p''^2-M^2)}
\ee
 which gives
\be
 \delta \tilde F^{\rm (el)}_{\rm box}(t,\nu) = 
  \frac{i \alpha t}{4\pi^3} (\nu - t)
   \int \left\{ 1 + \frac{2M^2+\nu-t}{\nu^2+t(4M^2-t)} t_p \right\}
    \bar F(t_1) \bar F(t_2) \frac{ d^4k''}{ k''^2(p''^2-M^2)}
\ee
 This quantity is marked with a tilde, because, due to the denominator
 of the second term in curly braces, it has unphysical poles % in $\nu$
 at $\nu = \pm \nu_0 = \pm \sqrt{-t(4M^2-t)}$.
 As described in Ref.~\cite{ourDisp}, we should subtract appropriate rational
 function of $\nu$ to obtain correct analytic behaviour of the amplitude.
 This is easily achieved with the help of
 Eq.~(B4) of Ref.~\cite{ourDisp}, yielding
\be \label{F_el}
 \delta F^{\rm (el)}_{\rm box}(t,\nu) = 
  \frac{i \alpha t}{4\pi^3} (\nu - t)
   \int \left\{ 1 + 
       \frac{(2M^2\!+\!\nu\!-\!t)t_p\!-\!(\nu\!-\!t)(p''^2\!-\!M^2)\!-\!(4M^2\!+\nu\!-\!t)k''^2}
       {\nu^2+t(4M^2-t)} \right\}
    \frac{ \bar F(t_1) \bar F(t_2) d^4k''}{k''^2(p''^2-M^2)}
\ee
 The subtracted quantity is a rational function of $\nu$, since
 the integrals $\int \bar F(t_1) \bar F(t_2) \frac{d^4k''}{p''^2-M^2}$
 and $\int \bar F(t_1) \bar F(t_2) \frac{d^4k''}{k''^2}$
 are independent of $\nu$;
 thus the subtraction does not introduce new cuts or violate unitarity.
 
 The full TPE amplitude is the sum of box and crossed-box amplitudes,
\be
 \delta F^{\rm (el)}(t,\nu) =
 \delta F^{\rm (el)}_{\rm box}(t,\nu) - \delta F^{\rm (el)}_{\rm box}(t,-\nu)
\ee
 The crossed box amplitude $ - \delta F^{\rm (el)}_{\rm box}(t,-\nu)$
 has no imaginary part in the $s$-channel ($\nu>0$),
 but provides correct imaginary part in the $u$-channel ($\nu<0$).
 
 The elastic part of the TPE amplitude (\ref{F_el}) is infra-red divergent.
 To obtain physically meaningful finite result,
 standard Mo\&Tsai contribution \cite{Tsai} is usually subtracted,
 and we also follow this way. % we do the same way.
 It can be shown that for $m \approx 0$ Mo\&Tsai contribution is equal to
\be
 \delta F^{\rm (MT)}_{\rm box}(t,\nu) = \frac{\alpha}{\pi} F(t)
   \left\{   2\ln\frac{\lambda}{M}\ln\frac{\nu-t}{2mM}
      + \ln^2\frac{m}{M}
      - \ln^2\frac{\nu-t}{2M^2}
      - \mathop{\rm Li_2}\left(1-\frac{\nu-t}{2M^2}\right)
   \right\}
\ee
 where $\lambda$ is infinitely small photon mass,
 introduced to regulate the divergence.
 The logarithmic dependence on $m$ disappears
 when adding box and crossed box amplitudes.
 
\subsection{$\rho$-meson inelastic contribution}

 This contribution arises from the $e\rho$ intermediate state
 in the r.h.s. of Eq.(\ref{uni}).
 To calculate it we need $\rho\pi\gamma^*$ vertex,
 which can be written in general form as
\be \label{rhoVert}
 \CM = \sqrt{4\pi\alpha} \frac{2M_\rho}{M_\rho^2-M^2}
 g(q^2) \ve^{\mu\nu\sigma\tau} e_\mu q_\nu p_\sigma v_\tau
\ee
 where $M_\rho$ is $\rho$-meson mass,
 $q$ and $p$ are photon and pion momenta,
 $e$ and $v$ are photon and $\rho$-meson polarizations, respectively,
 and $g(q^2)$ is dimensionless form factor.
 Its normalization is established from the $\rho \to \pi\gamma$ decay width
\be
 \Gamma_{\rho \to \pi\gamma} =
  \alpha |g(0)|^2 \frac{M_\rho^2-M^2}{6M_\rho}
\ee
 Using the latest value $\Gamma_{\rho \to \pi\gamma} = 68{\rm\ keV}$
 \cite{PDG}, we obtain $g(0) = 0.272$.
 For $q^2$ dependence we use simple vector-dominance-inspired form
\be
 g(q^2) = g(0) \frac{M_\omega^2}{M_\omega^2-q^2}
\ee
 where $M_\omega=0.872\GeV$.
 With $\rho\pi\gamma^*$ vertex (\ref{rhoVert}) we obtain
 the contribution to the imaginary part of the amplitude
\begin{eqnarray}
 \bar u'\gamma^\mu u \, (p+p')_\mu \, \Im \delta F^{(\rho)} (t,\nu) & = &
 \frac{\alpha t}{2\pi}
 \left( \frac{2M_\rho}{M_\rho^2-M^2} \right)^2
  \int \bar u'\gamma_{\mu'} \hat k'' \gamma_\mu u \times\\
 && \times  \ve^{\mu'\nu'\sigma'\tau} p'_{\nu'} p''_{\sigma'}
      \ve^{\mu\nu\sigma\tau} p_\nu p''_\sigma   
    \, \frac{g(t_1) g(t_2)}{t_1 t_2}
    \theta(p''_0) \delta(p''^2-M_\rho^2) \theta(k''_0) \delta(k''^2) d^4k'' \nonumber
\end{eqnarray}
% Following the same procedure as for the elastic contribution, we obtain
 The real part reconstruction procedure is the same as for the elastic
 contribution, except that the unphysical poles are subtracted
 with the help of the identity
\be
 \int d^4 p'' f(p'') \left. \left\{ 
      \frac{s t_p - (s-M^2)\Delta M^2}{k''^2(p''^2-M_\rho^2)}
      - \frac{s-M^2}{k''^2} - \frac{s+M^2}{p''^2-M_\rho^2} \right\}
   \right|_{\nu=\nu_0} = 0
\ee
 (where $\Delta M^2 = M_\rho^2 - M^2$),
 which can be obtained from Eq.(B4) of Ref.~\cite{ourDisp} putting
 $f(p'') \to f(p'')(p''^2-M^2)/(p''^2-M_\rho^2)$.
 
 The final result is
\begin{eqnarray}
 \delta F^{(\rho)}_{\rm box}(t,\nu) & = &
 -\frac{i \alpha t}{32\pi^2}
 \left( \frac{2M_\rho}{M_\rho^2-M^2} \right)^2
  \int \left\{ 
          \frac{2M^2 t_p - (\nu + t)(p''^2-M^2) - (4M^2-\nu-t) k''^2}{\nu^2+t(4M^2-t)}
            [t^2 + 2t\Delta M^2] + \right. \nonumber \\
&&      + \frac{(\nu-t)t_p + 2t(p''^2-M^2) - 2\nu k''^2}{\nu^2+t(4M^2-t)}
            [t^2 + t_1 t_2 - (t_p+2t)M^2 + (t-t_p) \Delta M^2 + \Delta M^4] + \nonumber \\
&&\left.+  2t_1 t_2 + t\Delta M^2 + t_p (\nu/2 - t - \Delta M^2) \vphantom{\int}\right\}
    \, \frac{g(t_1) g(t_2) d^4k''}{t_1 t_2 k''^2 (p''^2-M_\rho^2)}
\end{eqnarray}

\subsection{$b_1$ meson inelastic contribution}

Next allowed intermediate state with
sufficiently large $\pi\gamma$ branching ratio is $b_1(1235)$.
In general case, $b_1 \to \pi\gamma^*$ transition amplitude
depends on two form factors:
\be
 \CM = \sqrt{4\pi\alpha} \frac{2M_b}{M_b^2-M^2} 
 \left[ g_1(q^2) (q^\mu p^\nu - g^{\mu\nu} pq)
      + g_2(q^2) (q^2 g^{\mu\nu}- q_\mu q_\nu)
 \right] v_\mu e_\nu
\ee
Only $g_1$ contributes to $b_1 \to \pi\gamma$ decay width:
\be \label{b_Width}
 \Gamma_{b_1 \to \pi\gamma} = \alpha |g_1(0)|^2 \frac{M_b^2-M^2}{6M_b}
\ee
Thus we neglect second form factor and assume
\be
 g_1(q^2) = g_1(0) \frac{M_\omega^2}{M_\omega^2-q^2}, \qquad
 g_2(q^2) = 0
\ee
where $g_1(0)=0.40$ is obtained from Eq.~(\ref{b_Width}) and PDG value
$\Gamma_{b_1 \to \pi\gamma} = 230{\rm\ keV}$.

Further procedure is analogous to previous cases,
and the result is
\begin{eqnarray}
 \delta F^{(b_1)}_{\rm box}(t,\nu) & = &
 -\frac{i \alpha t}{32\pi^2}
 \left( \frac{2M_b}{M_b^2-M^2} \right)^2
  \int \left\{ 
          \frac{2M^2 t_p - (\nu + t)(p''^2-M^2) - (4M^2-\nu-t) k''^2}{\nu^2+t(4M^2-t)}
            [t^2 + 2t\Delta M^2] + \right. \nonumber \\
&&      + \frac{(\nu-t)t_p + 2t(p''^2-M^2) - 2\nu k''^2}{\nu^2+t(4M^2-t)}
            [t^2 + t_1 t_2 + t_p M^2 + (3t+t_p) \Delta M^2 + \Delta M^4] + \nonumber \\
&&\left.+  2t_1 t_2 + t\Delta M^2 + t_p (\nu/2 - t - \Delta M^2) \vphantom{\int}\right\}
    \, \frac{g_1(t_1) g_1(t_2) d^4k''}{t_1 t_2 k''^2(p''^2-M_b^2)}
\end{eqnarray}

\section{Results}

  In the numerical calculation we use monopole parameterization
  of pion form factor: $F(Q^2) = \Lambda^2/(Q^2+\Lambda^2)$,
  where $\Lambda = 0.719\GeV$ is chosen
  so as to reproduce measured charge radius of the pion.
  Its value as well as masses of all particles were taken from Ref.~\cite{PDG}.
  In what follows $\eps = \frac{\nu^2 - Q^2(4M^2+Q^2)}{\nu^2 + Q^2(4M^2+Q^2)}$
  is virtual photon polarization parameter.
    
\begin{figure}
 \hfill
  \includegraphics[width=0.48\textwidth]{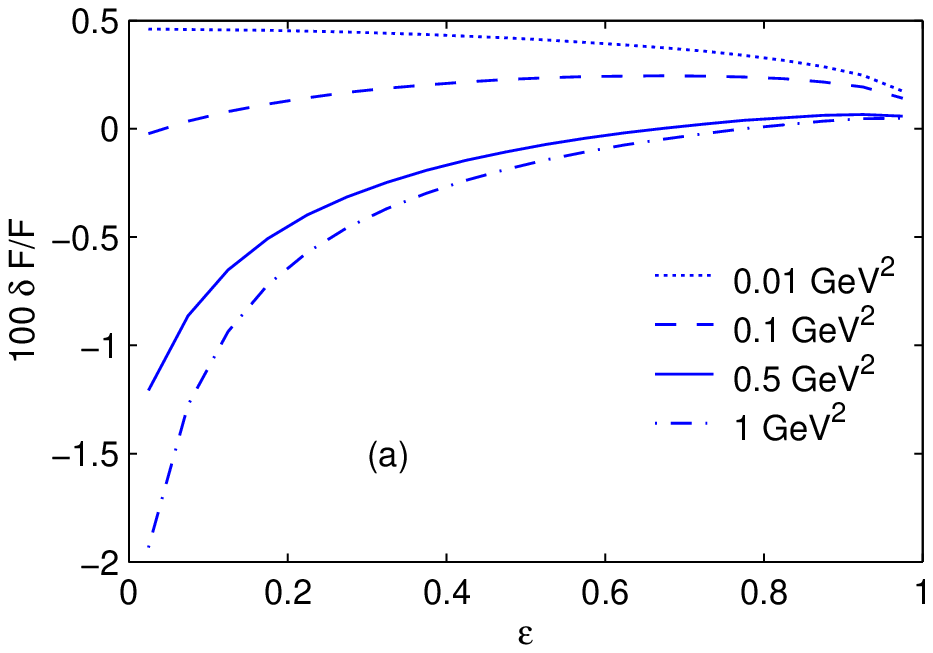}
 \hfill
  \includegraphics[width=0.48\textwidth]{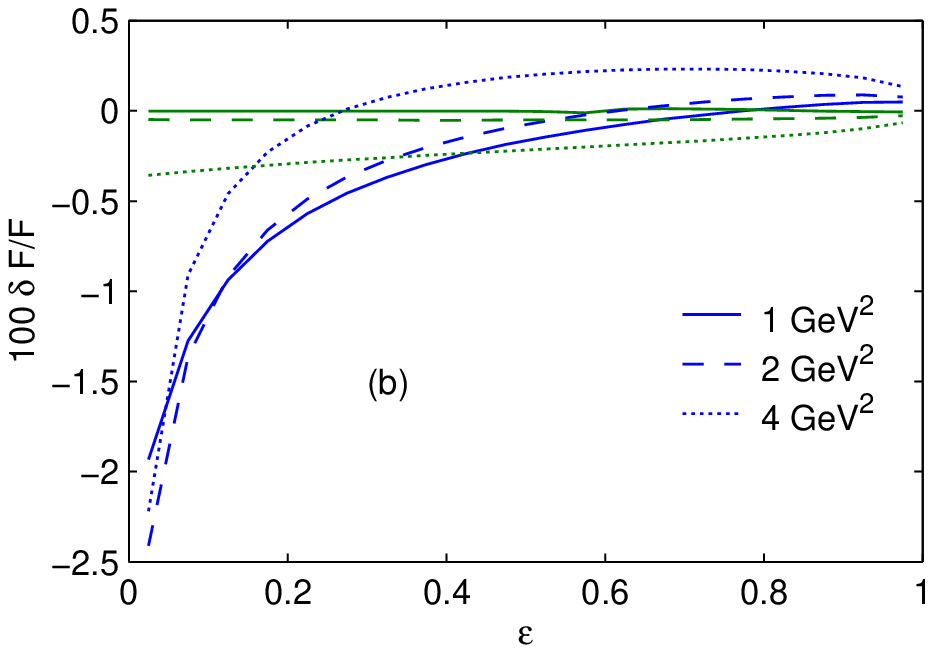}
 \hfill
 \caption{TPE amplitude as a function of $\eps$ at fixed $Q^2$,
   indicated on the figure.
   Elastic (blue) and inelastic (green) contributions,
   the former with Mo\&Tsai infra-red divergent contribution subtracted.
   On the left the inelastic contribution is negligibly small.
   } \label{fig:el}
\end{figure}

  The calculated elastic part of TPE amplitude (Fig.~\ref{fig:el}, blue curves)
  agrees well with the results presented in Ref.~\cite{piTPE2}.
  The coincidence is, most likely, accidental:
  we know that for the electron-proton scattering these two approaches
  give different analytical results \cite{ourDisp}.

\begin{figure}
 \hfill
  \includegraphics[width=0.48\textwidth]{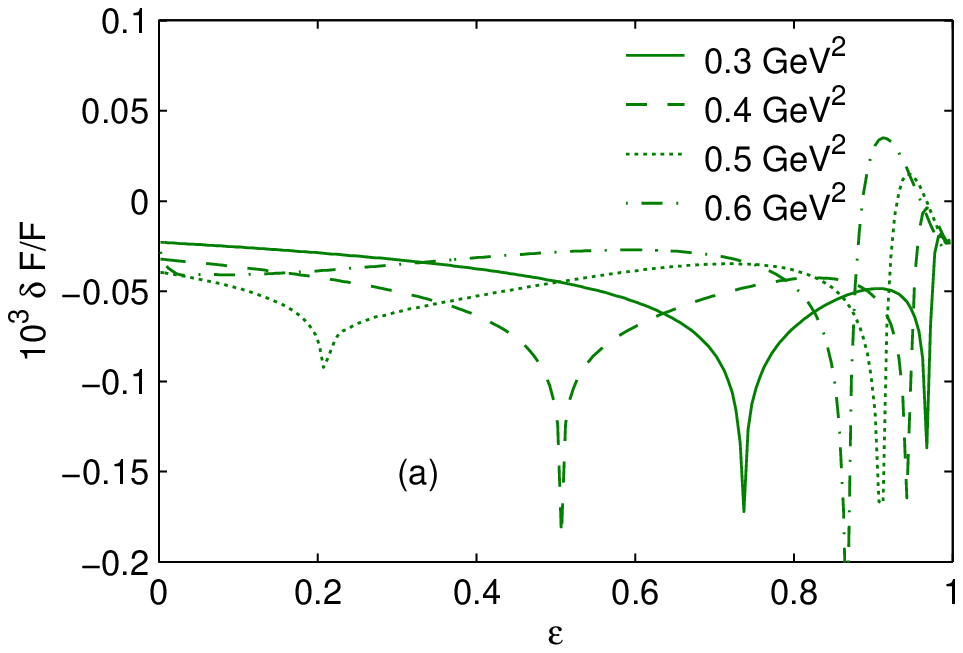}
 \hfill
  \includegraphics[width=0.48\textwidth]{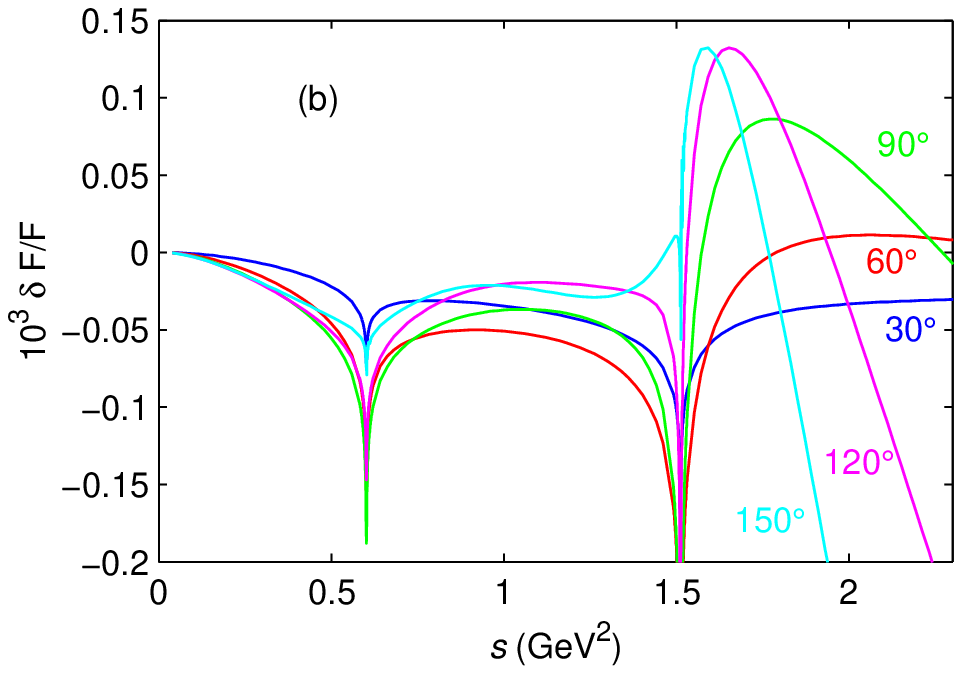}
 \hfill
 \caption{Inelastic contribution ($\rho + b_1$)
   as a function of $\eps$ at fixed $Q^2$ (a) and
   as a function of $s$ at fixed c.m. scattering angle (b).}\label{fig:inel}
\end{figure}

  The inelastic part of TPE amplitude,
  calculated with the inclusion of $\rho$
  and $b_1$ meson contributions, is shown in Fig.~\ref{fig:inel}.
More detailed numerical study reveal that it diverges logarithmically
at the thresholds $s=M_\rho^2$ and $s=M_b^2$.
This seems to be a consequence of neglecting respective meson widths.
If one takes finite width into account, the curves become "smeared" 
and the divergence should disappear.
  In comparison with the elastic one, the inelastic contribution
  is almost negligible, except at very high $Q^2$ (Fig.~\ref{fig:el}, right).
  However at high $Q^2$ our scheme for inelastic contribution
  calculation becomes doubtful.
  Indeed, we begin to exploit such an ambiguous thing as
  "the contribution of a resonance away from the resonance",
  and trust that it is main contribution.
  The inclusion of heavier resonances is likely needed.
  The contribution of non-resonant multi-particle states
  also may be significant, and needs to be estimated somehow.
  Therefore we think that at present the magnitude
  of the inelastic contribution at high $Q^2$ is not reliably known.
  
Both Refs.~\cite{piTPE1,piTPE2} and our work find that the elastic part
of TPE amplitude at high $Q^2$ sharply grows at backward angles
(i.e. near $\eps = 0$).
 The inelastic contribution has similar tendency,
 as one can infer from Fig.~\ref{fig:inel}(b).
Note that the amplitude does not diverge, it remains finite at $\eps = 0$.
In our approach this holds automatically,
since $\eps = 0$ is the physical region boundary,
corresponding to $\nu = Q\sqrt{4M^2+Q^2}$, whereas
the amplitude is constructed to be finite at this point
(\cite{ourDisp}, Sec.\ref{Sec:el}).

The explanation of this phenomenon is quite simple.
The full amplitude is the sum of box and x-box amplitudes,
and each of them has a singularity (a branching point)
%at $s=(M+m)^2$ or $u=(M+m)^2$, respectively.
%%at $\nu = \pm(t+4Mm)$.
%Though both singularities lie in the unphysical region,
%the $u$-channel singularity $u=(M+m)^2$ corresponds to
%$\eps = \frac{2t(M+m)^2}{t^2 + 16 M^2 m^2 - 2t(M-m)^2}$;
at $s=M^2$ or $u=M^2$, respectively (neglecting the electron mass).
Though both singularities lie in the unphysical region,
the $u$-channel singularity $u=M^2$ corresponds to
$\eps = -(1+Q^2/2M^2)^{-1}$;
for $Q^2 \gg M^2$ this is very close to $\eps=0$,
explaining the rapid amplitude growth near this point.
\begin{figure}
 \includegraphics[width=0.48\textwidth]{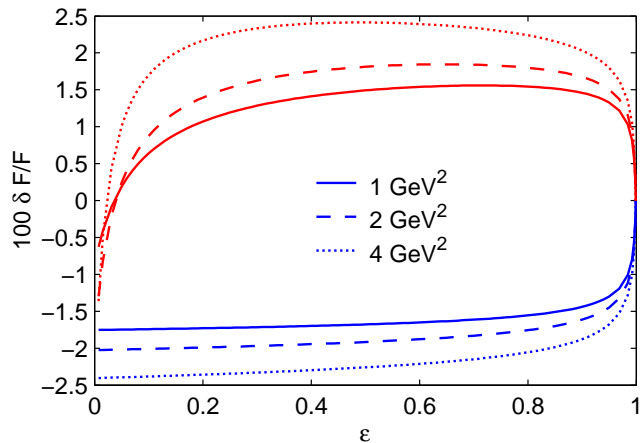}
 \caption{Box (blue, lower curves) and x-box (red, upper curves)
    parts of the elastic contribution.
  % Curve types same as in Fig.~\ref{fig:el}
   } \label{fig:su}
\end{figure}
A good illustration is Fig.~\ref{fig:su}, where box and x-box amplitudes
are plotted separately
(a proper constant was subtracted to make the amplitudes vanish at $\eps\to 1$).
Looking at the TPE amplitudes for the electron-proton scattering \cite{ourBox},
one can see a similar effect, which is just less pronounced,
because the proton mass is much higher.

\section{Conclusion}

 We have calculated TPE amplitude for the elastic electron-pion scattering
 in the dispersion approach,
 including both elastic and inelastic contributions.
 For the latter we take into account $\rho$ and $b_1(1235)$ mesons
 as intermediate states.
% Following common practice \cite{BMTres}, the latter
% was modelled as a sum of resonance contributions,
% namely, $\rho$ and $b_1(1235)$ meson contributions.

 We find that at not-so-high $Q^2$ (up to $2\GeV^2$)
 the elastic contribution is dominant, as in electron-proton scattering.
 At higher $Q^2$ the inelastic contribution is not negligible,
 but still smaller than the elastic one.
 However we believe that the former should be estimated more carefully,
 and no conclusion can be drawn at the moment.

 We also explain the behaviour of the amplitude at backward angles
 and $Q^2 \gg M^2$. As $Q^2$ increases,
 the $u$-channel threshold singularity approaches physical region boundary
 $\eps=0$ ($\theta=180^\circ$), resulting in sharp amplitude growth.

\end{document}